\documentclass[cits]{PoS}

\usepackage{feynmp}
\usepackage{stmaryrd}
\ifpdf
\fi
\usepackage{mathbbol}
\usepackage{tabularx}
\usepackage{multirow}
\usepackage{booktabs}
\usepackage{subfigure}
\usepackage[T1]{fontenc}
\usepackage[utf8]{inputenc}
\title{Static-static-light baryonic potentials}

\ShortTitle{Static-static-light baryonic potentials}

\author{\speaker{Johannes Najjar}, Gunnar Bali\\%
Institut f\"ur Theoretische Physik, Universit\"at Regensburg\\
93040 Regensburg, Germany\\
E-mail:\\ \email{johannes.najjar@physik.uni-regensburg.de},
\email{gunnar.gali@physik.uni-regensburg.de}}

\abstract{We determine doubly heavy baryonic
potentials as a function of the distance between the
two static sources, coupled to a light relativistic quark, for
different quantum numbers. We use the variational method to compute
the ground state and the first two excitations.
These can be used as an input to nonrelativistic models or to
NRQCD calculations of properties of doubly
heavy baryons. We compare our findings with a factorization model.
We employ all-to-all propagator methods, improved by an additional
hopping parameter expansion and Wuppertal smearing on
$N_f = 2$ QCDSF configurations.}

\FullConference{The XXVII International Symposium on Lattice Field Theory\\
July 26-31, 2009\\
Peking University, Beijing, China}

\begin{document}
\section{Introduction}
\begin{figure}
\centering
\subfigure{\includegraphics[width=0.24\textwidth]{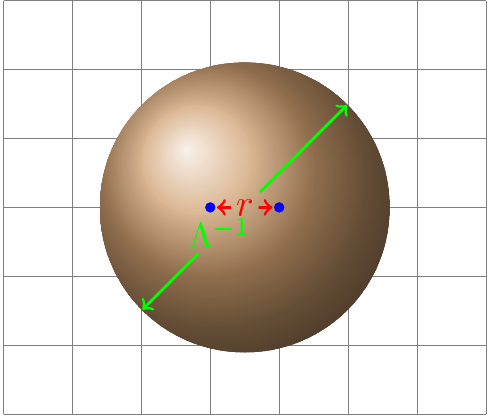}\label{fig:StaticStaticLightPicSavage_Wise_HQET}}
\subfigure{\includegraphics[width=0.24\textwidth]{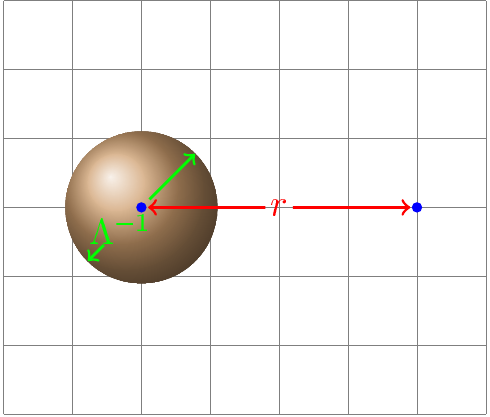}\label{fig:StaticStaticLightPic_LightOnStatic}}
\subfigure{\includegraphics[width=0.24\textwidth]{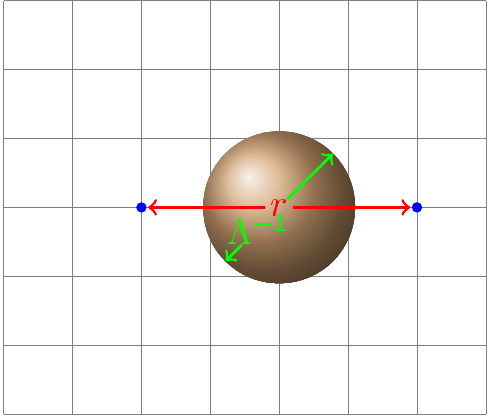}\label{fig:StaticStaticLightPic_LightInMiddle}}
\caption{The different models for the static-static-light baryon.} \label{fig:StaticStaticLightPic}
\end{figure}

The coupling of quarks inside a doubly heavy baryon is debated
and there are various models on the market like $\Delta$- or $Y$-flux
tubes between all three quarks or the formation of a diquark which couples
to the third quark.
The ground state doubly heavy baryon potential
has been studied previously on the lattice \cite{Suganuma}.
Ref.~\cite{Sava_White} suggests that a diquark
formation of the two heavy quarks is dynamically favored and that
the light quark perceives this as an almost pointlike object.
Therefore a $QQq$ baryon with two heavy quarks is related to a
heavy-light $\overline{Q}q$ meson~\cite{Sava_Wise}. In the static limit
where the spin of the heavy quark completely decouples,
the main
difference between these two systems is that the $QQ$ diquark
is spatially extended.  
In this work we check up to what $QQ$ separations this scenario holds.
In figure \ref{fig:StaticStaticLightPic} we depict different
situations. We ascribe a phenomenological
Compton wavelength $\Lambda\sim1/m_q$ to a
light quark of mass $m_q$. For $r\ll\Lambda^{-1}$ the light quark
 cannot
resolve the two static quarks (left picture).
When either the light quark mass or the static quark distance $r$ is increased,
the light quark can discriminate between the static color sources and
can ``choose'' its preferred localization. Symmetry considerations
suggest the scenarios
where the light quark sits on top of one of the static quarks
or in the middle between them to be of particular interest.
The first case corresponds to $Qq$ diquark formation while in the
second case no diquarks are formed.
Unlike in the physical situation of finite heavy quark masses where
the average distance $\langle r\rangle$ is dynamically determined,
in the static limit we can change $r$ to interpolate between the
$QQ$-diquark picture at small $r$ and the other more involved
scenarios. At $r=0$ the spectra of the $QQq$ and
$\overline{Q}q$ situations are (numerically and analytically) identical.
As $r$ is increased the energies will increase. For instance,
$V_{QQ}=V_{\overline{Q}Q}/2$, up to a constant, at least to the first
two orders of perturbation theory and in the string picture.
We denote mass levels of the mesonic  $\overline{Q}q$ system
by $m_{\overline{Q}q}$. Thus, in the $QQ$-diquark picture, that at least should
hold at $r<\Lambda^{-1}$, we expect the factorization,
\begin{equation}
V_{QQq}(r)\approx m_{\overline{Q}q} +\frac{1}{2} V_{\overline{Q}Q}(r)\,,
\end{equation}
which can graphically be depicted as,
\begin{equation}
\left .
\exp\left(
\;-\;\;
\parbox{.07\textwidth}{
\begin{fmffile}{factorization_middle}
\begin{fmfgraph*}(30,30)
\fmfstraight
\fmfleft{lu,lo}	
\fmfright{ru,ro}
\fmf{fermion}{lu,lo}
\fmf{fermion}{ru,ro}
\fmf{plain}{ro,om}
\fmf{plain}{ru,um}
\fmf{plain}{lo,om}
\fmf{plain}{lu,um}
\fmf{wiggly,tension=0}{um,om}
\fmf{phantom_arrow,tension=0}{um,om}
\end{fmfgraph*}
\end{fmffile}}
\;\;\;
\right) 
\;
\propto
\exp\left(
\;-\;
\parbox{.025\textwidth}{
\begin{fmffile}{static_light}
\begin{fmfgraph*}(11,30)
\fmfstraight
\fmfkeep{static_light}
\fmfright{ru,ro}
\fmf{fermion}{ro,ru}
\fmf{wiggly,tension=0,left=0.5}{ru,ro}
\fmf{phantom_arrow,tension=0,left=0.5}{ru,ro}
\end{fmfgraph*}
\end{fmffile}}
\;\;-\;
\frac{1}{2}
\;\;
\parbox{.07\textwidth}{
\begin{fmffile}{Wilsonloop}
\begin{fmfgraph*}(30,30)
\fmfstraight
\fmfkeep{Wilsonloop}
\fmfleft{lu,lo}	
\fmfright{ru,ro}
\fmf{fermion}{lu,lo}
\fmf{fermion}{ro,ru}
\fmf{plain}{ro,om}
\fmf{plain}{ru,um}
\fmf{plain}{lo,om}
\fmf{plain}{lu,um}
\end{fmfgraph*}
\end{fmffile}}
\;\;\; 
\right )
\right \uparrow_T\,,
\end{equation}
where straight lines denote gauge transporters (including
the static propagators)
and wiggly lines stand for light quark propagators.
The $QQq$ potentials only depend on the distance and
a $D_{\infty h}'$ representation (see below). The light quark and gluon
degrees of freedom have been integrated out. 
One can use the resulting potentials as an input
to a Schr\"odinger equation, in the heavy quark limit,
and apply the formalism of NRQCD for corrections.

\section{Simulation details}
We use 68 QCDSF configurations \cite{Brommel2007}. The specifications
are displayed in table \ref{table:lattices}.
\begin{table}[h]
\begin{center}
\begin{tabular}{|l|l|l|l|l|l|l|l|}\hline
Lattice points & $\kappa_{val} $  &  $\kappa_{sea} $  & $\beta$&  $C_{SW}$ &  $m_{\pi}$ [GeV]  & $a$ [fm] & $L$ [fm]\\\hline
$16^3\times 32$  &  0.1355  &  0.1355  &  5.29  &  1.9192  &  0.783(11) & 0.084(1)  &  1.3\\\hline
\end{tabular}
\caption{Wilson-Clover $N_f=2$ lattices.}\label{table:lattices}
\end{center}
\end{table}
We use all-to-all propagators with 300 stochastic estimates
on each configuration. We apply the hopping parameter expansion
\cite{Thron,Sesam} to reduce the stochastic noise.
APE-smeared \cite{Ape1,Ape2} gauge links were used for the
spatial gauge connectors in the Wilsonloops and the link
covariant displacements. For the static quark propagator
we use temporal links that have been stout smeared \cite{Stout} once,
to reduce the static energy and thus to improve the signal to noise ratio.
The quark fields of the all-to-all propagators are Wuppertal
smeared \cite{Wuppertalsmear} using APE-smeared gauge transporters and
we applied the variational method \cite{Var1,Var2} to extract
the groundstate (GS), the first excitation and second excitations
(FE, SE). With this setting we follow the methods described in \cite{Sesam}.
We compute the baryon correlator with two static quarks in various geometries.
The light quark can be in the middle which we will refer to as $\boxbar$,
\begin{equation}
\begin{fmffile}{diagm}
\begin{fmfgraph*}(30,30)
\fmfstraight
\fmfleft{lu,lo}	
\fmfright{ru,ro}
\fmf{fermion}{lu,lo}
\fmf{fermion}{ru,ro}
\fmf{plain}{ro,om}
\fmf{plain}{ru,um}
\fmf{plain}{lo,om}
\fmf{plain}{lu,um}
\fmf{wiggly,tension=0}{um,om}
\fmf{phantom_arrow,tension=0}{um,om}
\end{fmfgraph*}
\end{fmffile},
\end{equation}
or the light quark can be placed
at one of the static quark positions, symmetric ($\oplus$) or anti-symmetric ($\ominus$),
\begin{equation}
\left( \;\;
\parbox{.07\textwidth}{
\begin{fmffile}{rechts}
\begin{fmfgraph*}(30,30)
\fmfstraight
\fmfkeep{rechts}
\fmfleft{lu,lo}	
\fmfright{ru,ro}
\fmf{fermion}{lu,lo}
\fmf{fermion}{ru,ro}
\fmf{plain}{ro,om}
\fmf{plain}{ru,um}
\fmf{plain}{lo,om}
\fmf{plain}{lu,um}
\fmf{wiggly,tension=0,left=0.5}{ru,ro}
\fmf{phantom_arrow,tension=0,left=0.5}{ru,ro}
\end{fmfgraph*}
\end{fmffile}}
\;\;\;+\;\;\;
\parbox{.07\textwidth}{
\begin{fmffile}{links}
\begin{fmfgraph*}(30,30)
\fmfstraight
\fmfkeep{links}
\fmfleft{lu,lo}	
\fmfright{ru,ro}
\fmf{fermion}{lu,lo}
\fmf{fermion}{ru,ro}
\fmf{plain}{ro,om}
\fmf{plain}{ru,um}
\fmf{plain}{lo,om}
\fmf{plain}{lu,um}
\fmf{wiggly,tension=0,right=0.5}{lu,lo}
\fmf{phantom_arrow,tension=0,right=0.5}{lu,lo}
\end{fmfgraph*}
\end{fmffile}}
\;\;\;
\right)
\pm
\left(\;\;\;
\parbox{.07\textwidth}{
\begin{fmffile}{diagr}
\begin{fmfgraph*}(30,30)
\fmfstraight
\fmfkeep{diagr}
\fmfleft{lu,lo}	
\fmfright{ru,ro}
\fmf{fermion}{lu,lo}
\fmf{fermion}{ru,ro}
\fmf{plain}{ro,om}
\fmf{plain}{ru,um}
\fmf{plain}{lo,om}
\fmf{plain}{lu,um}
\fmf{wiggly,tension=0}{ru,lo}
\fmf{phantom_arrow,tension=0}{ru,lo}
\end{fmfgraph*}
\end{fmffile}}
\;\;\;+\;\;\;
\parbox{.07\textwidth}{
\begin{fmffile}{diagl}
\begin{fmfgraph*}(30,30)
\fmfkeep{diagl}
\fmfstraight
\fmfleft{lu,lo}	
\fmfright{ru,ro}
\fmf{fermion}{lu,lo}
\fmf{fermion}{ru,ro}
\fmf{plain}{ro,om}
\fmf{plain}{ru,um}
\fmf{plain}{lo,om}
\fmf{plain}{lu,um}
\fmf{wiggly,tension=0}{lu,ro}
\fmf{phantom_arrow,tension=0}{lu,ro}
\end{fmfgraph*}
\end{fmffile}}
\;\;\;
\right).
\end{equation}
All these correlators are realized by using the appropriate gauge connectors 
$\mathcal{P}$ in the baryon interpolator
\begin{equation}
O(x,x',x'')^\alpha= \varepsilon_{abc}P_+^{\alpha \alpha'}   Q(x)^{\alpha'}_{a'} \mathcal{P}(x, x'')_{a'a} \left(Q(x')^{\beta}_{b'} \mathcal{P} (x', x'')_{b'b}\Gamma^{\beta\gamma}q(x'')^{\gamma}_{c}\right)\,, \label{eqn:Annihiliator}
\end{equation}
where all coordinates $x$ are at the same time $t$, $P_+= \frac12(\mathbb1 + \gamma_4)$ is the projector to positive parity and $\Gamma$
is one of the operators from table \ref{Table:List_of_Operators}.
We denote the separation of the two heavy quarks by $r$. We
use the same operator $\Gamma$ for source and sink  so that the
correlator we compute is characterized by $r,\Gamma$,
the temporal extent of the correlator $T$ and the geometry
( $\boxbar,\oplus,\ominus$ ). For $r=0$,  $\boxbar$ and $\oplus$
are degenerate.

The \textit{octahedral group} $O_h$ is the cubic group of rotations on the lattice, with the addition of parity. Its irreducible representations describe point-particles at rest on a discrete lattice. The mapping between these and the continuum quantum numbers is not unique and so an lattice irreducible representation can correspond to multiple continuum $J$ states.\\
In our case, at $r>0$, this symmetry is broken to the cylindrical subgroup $D_{4h}\subset O_h$. In the continuum this corresponds to $D_{\infty h}\subset O(3)$.
 As we are interested in baryons we have to take a spinorial irreducible representation (irrep). We refer to these representations
with half integer
values of $J$ as $O_h'$ or $D_{\infty h}'$/$D_{4 h}'$.
\begin{table}[h]
\begin{center}
\begin{tabular}{|c|c|c|c|}\hline
& &$r=0$ &$r>0$ \\\hline
Wave & Operator & $O'(3)$, $O_h'$& $D_{\infty h}'$, $D_{4 h}'$ \\ \hline
$S$ & $\gamma_5$& $\frac{1}{2}^+$, $G_1^+$ & $\frac{1}{2}_g$, $G_{1g}$\\
\hline
$P_-$ & $\mathbb1$ & $\frac{1}{2}^-$, $G_1^-$ & $\frac{1}{2}_u$, $G_{1u}$\\
\hline 
\multirow{2}{*}{$P_+$} &\multirow{2}{*}{ $\gamma_1\Delta_1-\gamma_2\Delta_2$ $\oplus$ cyclic} & \multirow{2}{*}{$\frac{3}{2}^-$, $H^-$} & $\frac{3}{2}_u$ $\parallel$, $G_{2u}$  \\
&&&$\frac{1}{2}_u$ $\perp$, $G_{1u}$  \\
\hline
\multirow{2}{*}{$D_-$} &\multirow{2}{*}{$\gamma_5(\gamma_1\Delta_1-\gamma_2\Delta_2)$ $\oplus$ cyclic  } & \multirow{2}{*}{$\frac{3}{2}^+$, $H^+$} & $\frac{3}{2}_g$ $\parallel$, $G_{2g}$  \\
&&&$\frac{1}{2}_g$ $\perp$, $G_{1g}$ \\\hline
$D_+$ & $\gamma_1\Delta_2\Delta_3+\gamma_2\Delta_3\Delta_1+\gamma_3\Delta_1\Delta_2$ & $\frac{5}{2}^+$, $G_1^+$ & $\frac{1}{2}_g$, $G_{1g}$\\ 
\hline
$F_-$ &$\gamma_5(\gamma_1\Delta_2\Delta_3+\gamma_2\Delta_3\Delta_1+\gamma_3\Delta_1\Delta_2)$  &$\frac{5}{2}^-$, $G_1^-$ & $\frac{1}{2}_u$/$\frac{5}{2}_u$, $G_{1u}$
\\\hline
\end{tabular}
\caption{List of operators and representations.}\label{Table:List_of_Operators}
\end{center}
\end{table}
In table \ref{Table:List_of_Operators} we list the operators that we have used to calculate different correlators and their corresponding quantum numbers.
We denote the link covariant displacement as,
\begin{equation}
\Delta_\mu \eta(x) = U_\mu(x) \eta(x+a\mu)  - U_{-\mu}(x)\eta(x-a\mu) \quad \mathrm{where} \quad U_{-\mu}(x)= U^\dagger_{\mu}(x-\mu) \;.
\end{equation}
We have the correspondence 
\begin{equation}
G_1 \leftarrow \frac{1}{2}, \frac{5}{2}, \cdots \quad \mathrm{and}\quad G_2 \leftarrow \frac{3}{2}, \frac{7}{2}, \cdots
\end{equation}
between the $D_{4h}'$ lattice and the $D_{\infty h}'$ continuum quantum numbers. For $r>0$ some $O_h'/O(3)'$ irreps will split up
into two or more $D_{4h}'/D_{\infty h}'$ irreps. For instance,
the operators corresponding to $H^{\pm}$ split up, depending on the direction,
 relative to the interquark axis: for the axis pointing into the $\hat{3}$-direction,
we call the operator $\gamma_1\Delta_1-\gamma_2\Delta_2$ ``parallel'' ($\parallel$)
and the other combinations ``perpendicular'' ($\perp$), where $\perp$ corresponds to the state of
lower angular momentum ($G_1/\frac12$) and $\parallel$ to the higher angular momentum  ($G_2/\frac32$), relative to
the $\hat{3}$-axis.

\section{Results}
\subsection{Spectrum}
\begin{figure}
\centering
\subfigure[$r=0$]{\label{fig:allops_regge}\includegraphics[width=0.475\textwidth]{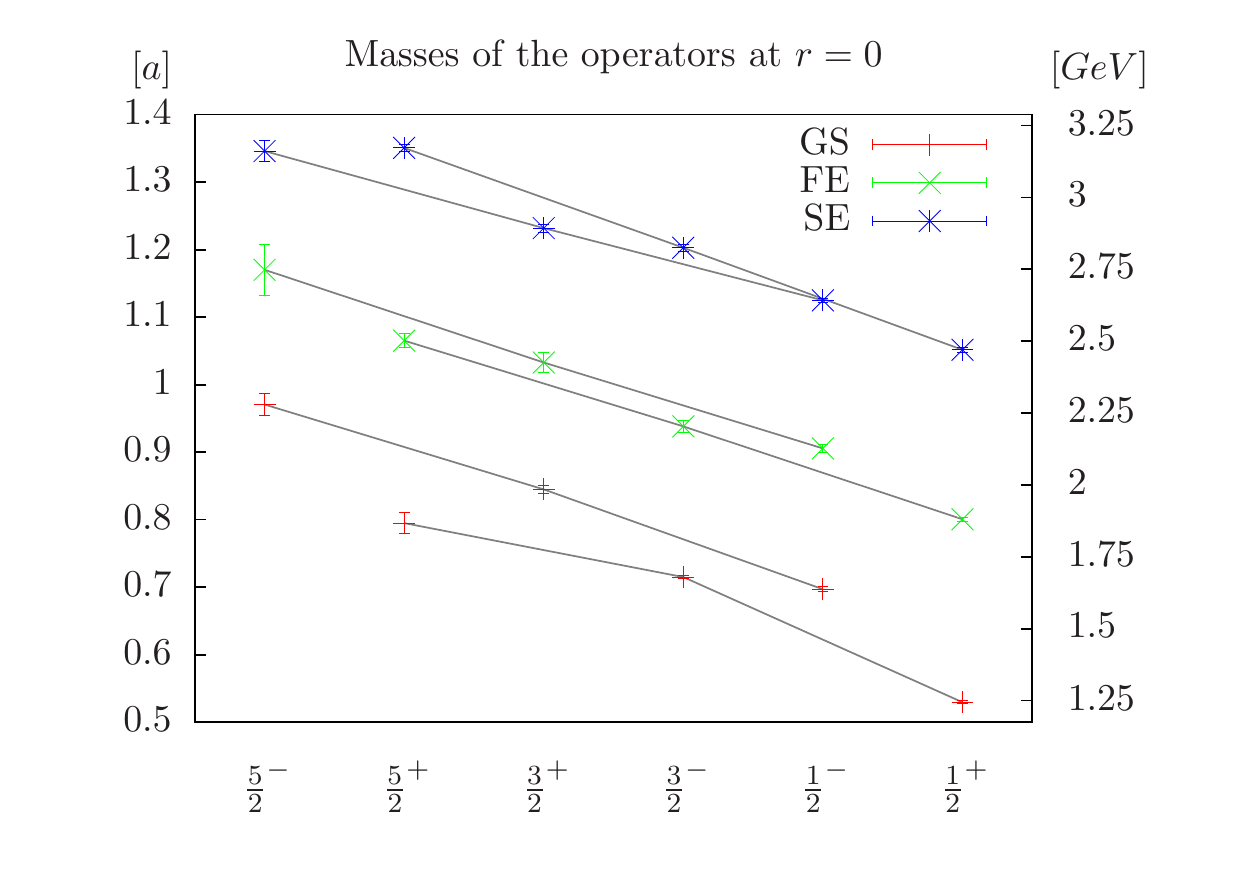}}                
\subfigure[$r=5a$]{\label{fig:allops_r5}\includegraphics[width=0.475\textwidth]{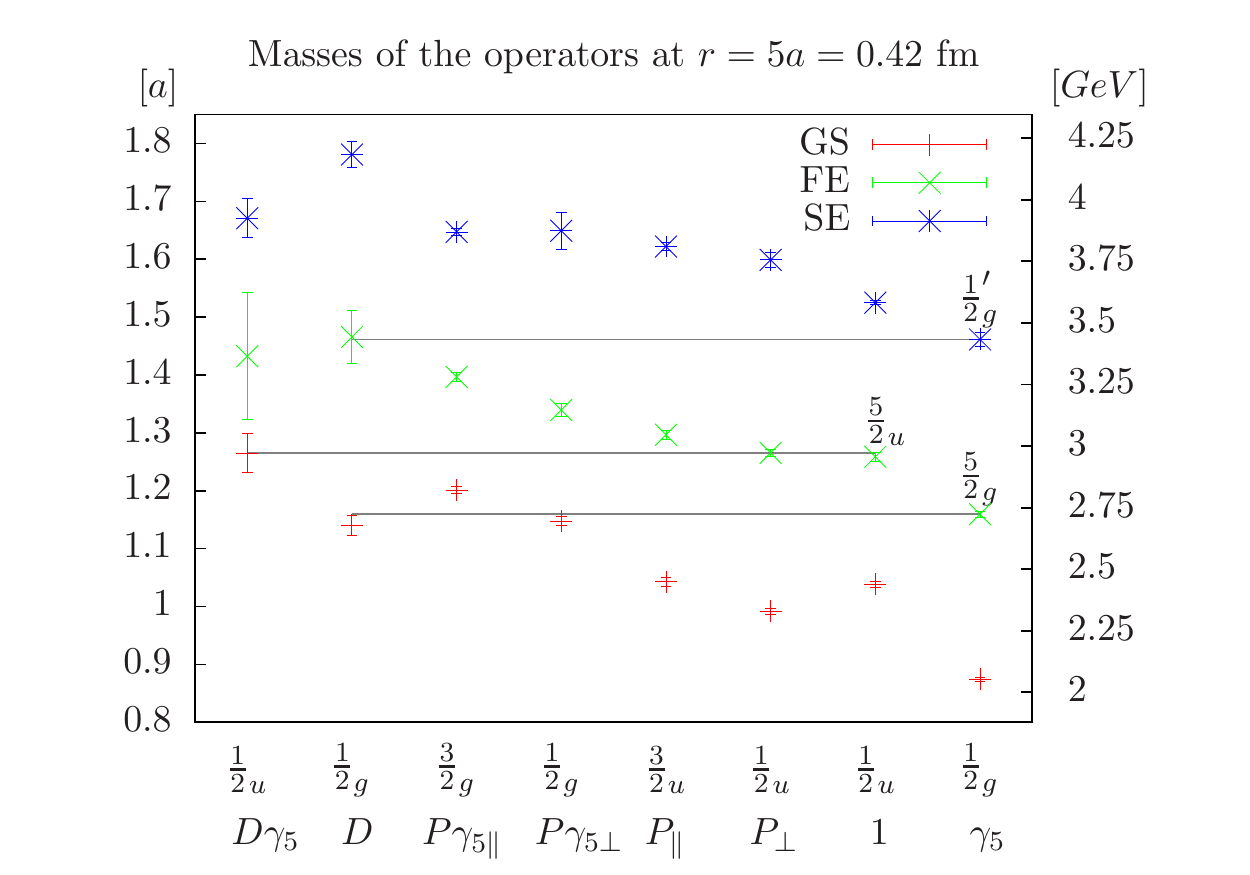}}
\caption{The different operators at $r=0$ and $r=5a$, with the lowest
possible continuum spin assignments.}
\label{fig:allops}
\end{figure}

The spectrum of our observables at $r=0$ is shown in
figure \ref{fig:allops_regge}. The Regge trajectories are denoted
by the grey lines and although they are not examined further
they help to group the operators. The spectrum extends over
a range of $2$ GeV and it gets even richer when one goes
to $r>0$ where the symmetry group $O_h'$ breaks down into $D_{4 h}'$.

The states created by the operators $P$ and $P \gamma_5$ split up
into two energy levels depending on whether or not the angular
momentum contains a projection onto the $QQ$ axis.
The spectrum at $r=5a$ is shown in figure \ref{fig:allops_r5}
and the grey lines in the plot show degeneracies of
some of the levels. For example, the excitation of $\mathbb{1}$ and
the ground state of $D\gamma_5$ are degenerate. Both
are in the $G_{1u}$ representation of $D_{4h}'$, corresponding
to the continuum $D_{\infty h}'$ quantum numbers $\frac12_u,\frac52_u,\ldots$.
The latter operator corresponds to
the $O(3)$ quantum numbers $\frac{5}{2}^-$ at $r=0$. Hence we
assign a continuum $\frac52_u$ spin to it and to the first radial
excitation of the $1$ operator.
Unfortunately the two ground states of $\frac{1}{2}_u$, i.e.\ $\mathbb{1}$
and $P \perp$ seem to be different and it looks
like $\frac{1}{2}_u \simeq \mathbb{1}$ is much closer
to $\frac{3}{2}_u \simeq P \parallel$. We hope to resolve this issue in the
near future.

\begin{figure}[ht]
\begin{minipage}[b]{0.475\linewidth}
\centering
\includegraphics[width=\textwidth]{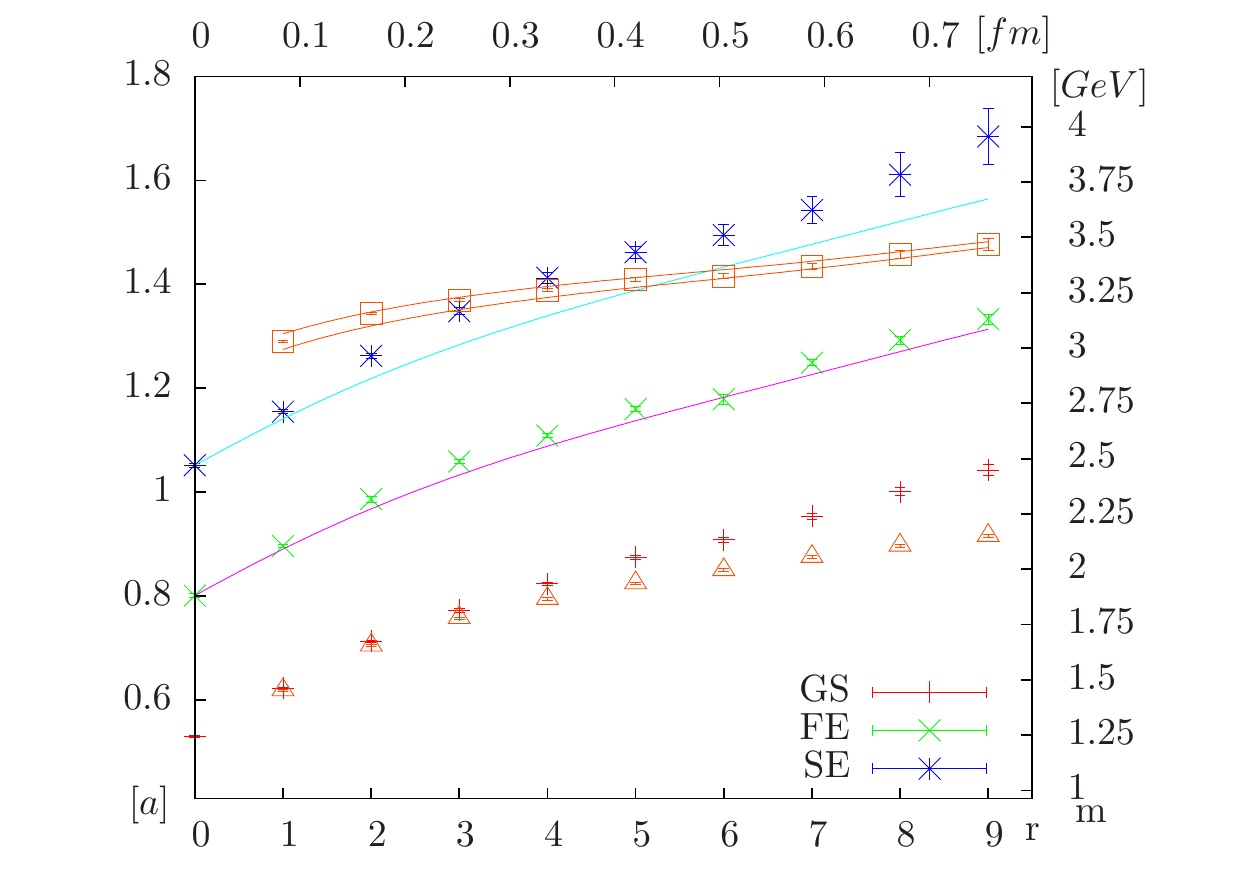}
\caption{Potential for $\frac{1}{2}_g \simeq \gamma_5 \oplus$.}\label{fig:Potential_g5}
\label{fig:figure1}
\end{minipage}
\hspace{0.5cm}
\begin{minipage}[b]{0.475\linewidth}
\centering
\includegraphics[width=\textwidth]{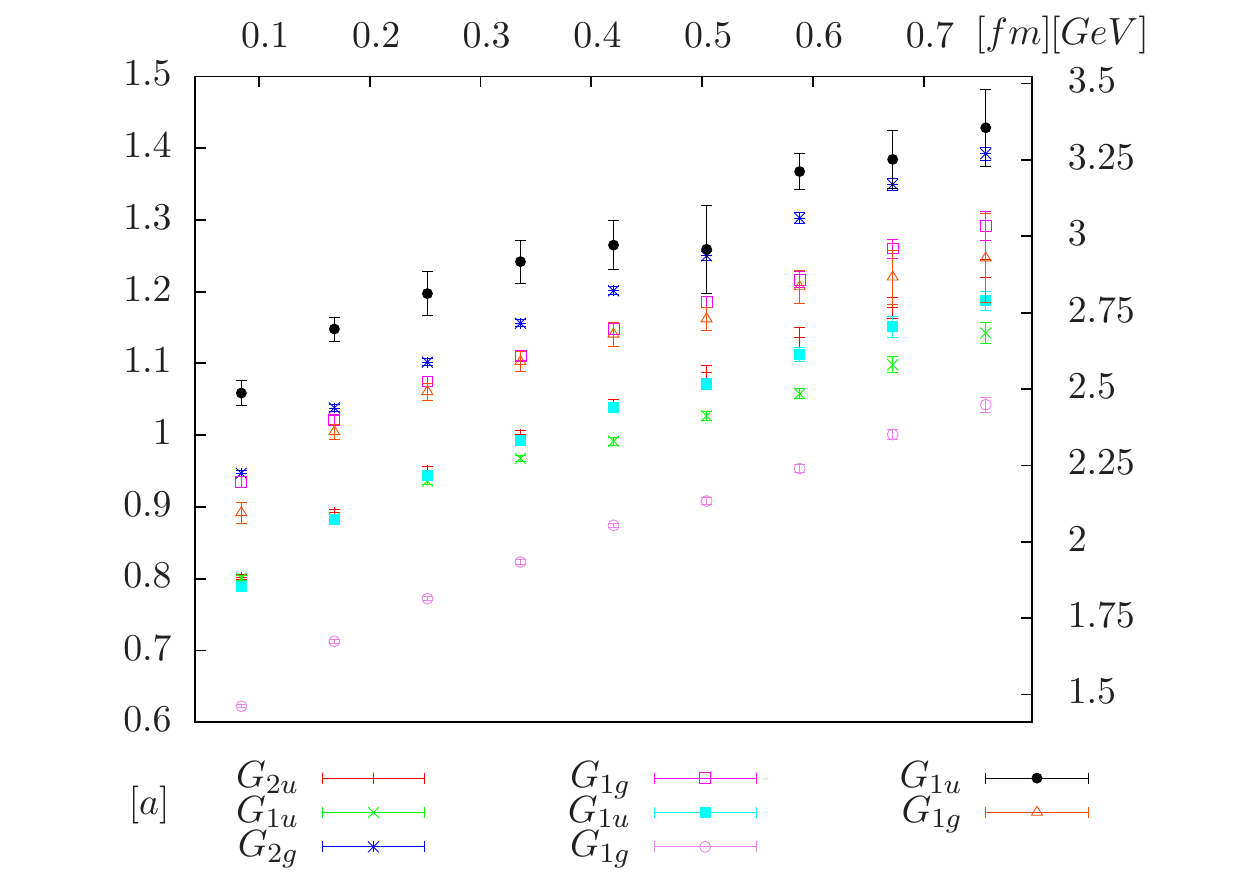}
\caption{Lowest lying potential for each operator.} \label{fig:allopp_split}
\end{minipage}
\end{figure}

\subsection{Potentials}

In figure \ref{fig:Potential_g5} we plot the results of $\frac{1}{2}_g \simeq \gamma_5 \oplus$ as red, green and blue symbols. 
The factorization model expectation 
for the ground state $m_{\overline{Q}q}+\frac{1}{2}V_{Q\overline{Q}}$ is
represented by the orange triangles. For $r>3a\approx 0.25$ fm the data lie
systematically above the expectation. 
The increase in energy with increasing $r$ originates from
the internal energy of the $QQ$ diquark which, in the presence of the
light quark, rises more steeply then expected. In figure \ref{fig:allopp_split} 
we see that this slope depends on the $D_{4h}'$ quantum numbers, such that
this failure of the factorization ansatz cannot be attributed to
a different functional form of the $QQ$ potential alone.

The pink and light blue lines are interpolations of the ground state points,
shifted by the first two static-light energy splittings. The
pink line describes the first excitation of the baryonic potential
very well. This suggests that 
this excitation is due to
the light quark, with very little effect from the gluonic
flux configuration.
The second excited state data lie somewhat above the light blue line.  
The most likely reason for this deviation is our inability to
reliably isolate this excitation in our three-dimensional
variational basis, so that we somewhat overestimate the masses.

At large $r$ one of the static quarks might
form a diquark with the light quark. In this case we should see
the transverse modes of the string connecting this $Qq$ diquark
with the remaining $Q$ in the excitation spectrum.
The Nambu-Goto string potential \cite{Arvis} suggests the following
functional form:
\begin{equation}
E_n(r)=\sigma_{\rm GS}r\sqrt{1+\left(2n-\frac{d-2}{12}\right)\frac{\pi}{\sigma_{\rm GS} r^2}} \;.
\end{equation}
In our case $d=4$ and $\sigma_{\rm GS}$ is the effective string tension
of the ground state determined by fitting to
\begin{equation}
V_{QQq}(r)=C_{\rm eff} + \sigma_{\rm eff}r - \frac{A_{\rm eff}}{r} \;. \label{eqn:Potential_model}
\end{equation}
The red band is $E_2-E_0+\mathrm{\rm GS}$, which is the first candidate for
an excitation because $n=1$ would affect the spin of the system.
This curve is above our first radial excitation and
its shape is very different from the second excitation.
The light quark excitation energies are smaller than
those required for transverse excitations of the flux tube.  
The plots for the other observables look very similar.

\subsection{Wavefunctions}
We compute the RMS and the wavefunctions of the ground state and
the first two excitations as outlined in \cite{CEhmann}.
The light quark RMS for the ground state wavefunction created by the
operator $\gamma_5 $ is about $4.1a$. Within errors it is independent of
the geometry of the correlator. This distance is indeed reasonably close
to $r\approx 0.3$~fm, where the factorization ansatz starts to fail.
The shapes of the Coulomb gauge wavefunctions agree with the findings
of \cite{CEhmann} although this reference considered mesons
while we study the light quark distribution inside a $QQq$-baryon.
\begin{figure}[h]
\begin{center}%
\subfigure{\includegraphics[width=0.32\textwidth]{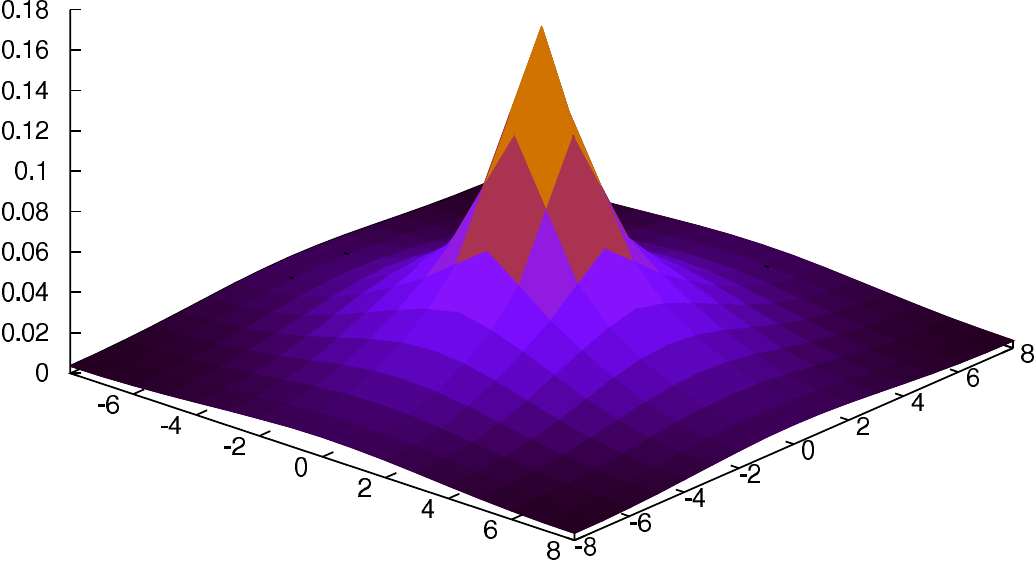}}%
\subfigure{\includegraphics[width=0.32\textwidth]{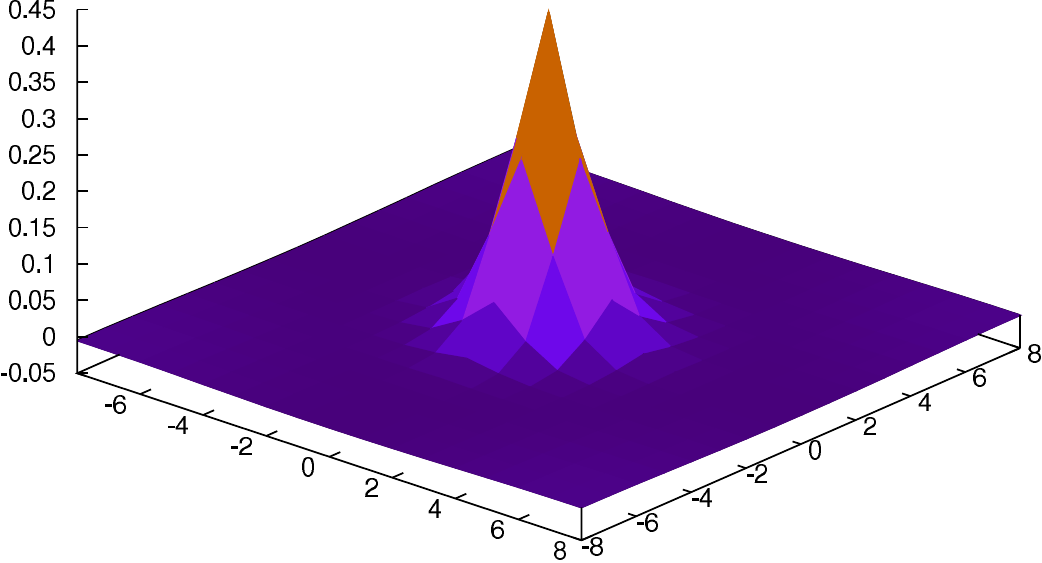}}%
\subfigure{\includegraphics[width=0.32\textwidth]{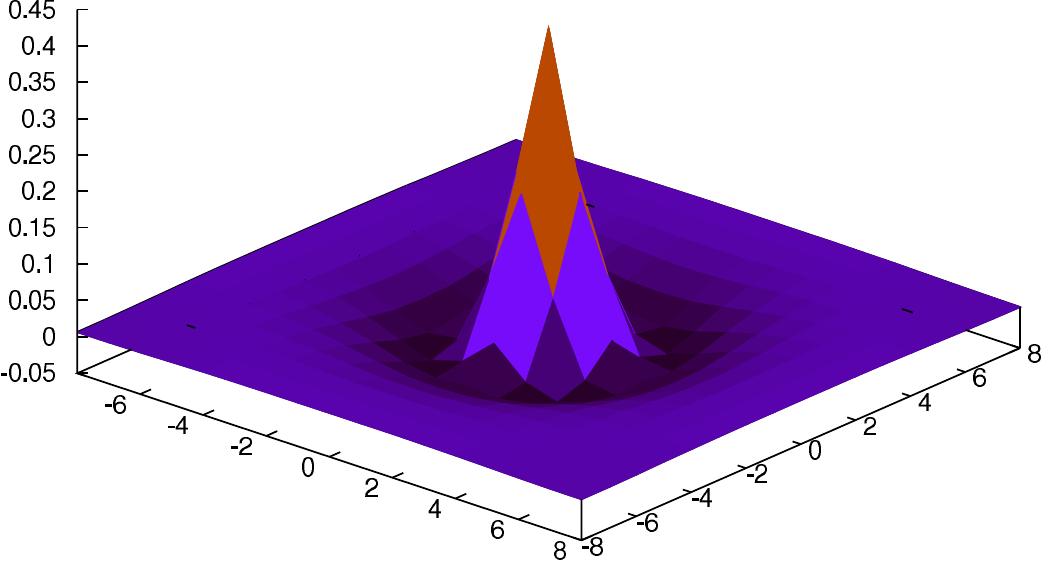}}%
\caption[Wavefunction of $\gamma_5 \oplus$]{Wavefunction of groundstate and first two excitations of $\gamma_5 \oplus$ for $r=4$.}\label{fig:WF_g5_Plus}
\end{center}
\end{figure}%
\section{Conclusion and Outlook}
The factorization model breaks down for separations greater than $0.3$ fm. The nature of the first excitation of a static-static-light baryon does not change with an increase of the separation in the two static quarks. The second excitation could not be resolved so well and we do not venture to explain its nature. The splittings induced by breaking the symmetry group $O_h$ into $D_{4h}$ are reflected in our correlators.
\\
It is a puzzle why the states of $\frac{1}{2}_u$ coming from the operators $\mathbb{1}$ and from $P_\perp$ do not agree but instead the $\mathbb{1}$ seems to be degenerate with the $\frac{3}{2}_u$ from $P_\parallel$. At the moment we are conducting a study on a bigger lattice ($24^3\times48$) at smaller quark masses ($\kappa=0.1362$) and we hope to clarify this point.

\acknowledgments
We thank the QCDSF Collaboration for making their configurations
available on the ILDG. We use the chroma package \cite{Chroma}
for our simulations. This work was supported by the DFG
Sonderforschungsbereich/Transregio~55 and BMBF grant 06RY257.

\end{document}